\documentclass{article}

\usepackage[preprint]{neurips_2022}

\usepackage[utf8]{inputenc} % allow utf-8 input
\usepackage[T1]{fontenc}    % use 8-bit T1 fonts
\usepackage{hyperref}       % hyperlinks
\usepackage{url}            % simple URL typesetting
\usepackage{booktabs}       % professional-quality tables
\usepackage{amsfonts}       % blackboard math symbols
\usepackage{nicefrac}       % compact symbols for 1/2, etc.
\usepackage{microtype}      % microtypography
\usepackage{xcolor}         % colors
\usepackage{colortbl}

\usepackage{natbib}
\setcitestyle{numbers,square}

\usepackage{amsmath}
\usepackage{svg}
\usepackage{algorithm}
\usepackage{algorithmic}
\usepackage{framed}
\usepackage{adjustbox}
\usepackage{multirow}
\usepackage{graphicx}
\usepackage{amsthm}
\usepackage{amssymb}
\theoremstyle{plain}
\newtheorem{theorem}{Theorem}%[section]

\theoremstyle{definition}
\newtheorem{definition}[theorem]{Definition}
\theoremstyle{remark}

\definecolor{cblue}{RGB}{218,232,252}
\definecolor{cgreen}{RGB}{213,232,212}
\definecolor{cyellow}{RGB}{255,242,204}
\definecolor{cred}{RGB}{248,206,204}
\definecolor{corange}{RGB}{255,230,204}
\definecolor{cgray}{gray}{.9}

\makeatletter
\def\@fnsymbol#1{\ensuremath{\ifcase#1\or \includegraphics[width=0.45cm]{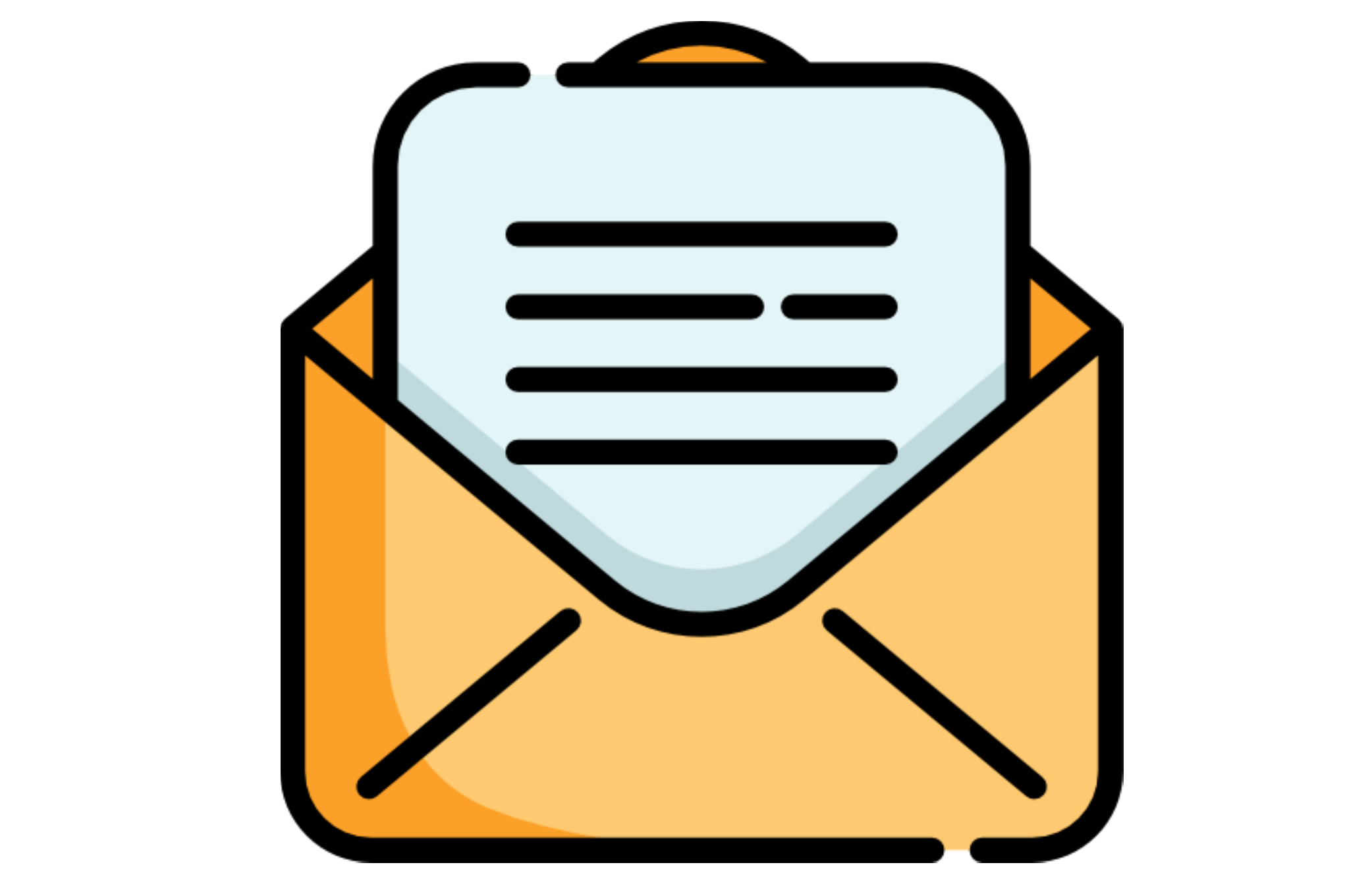} \or *\or %\dagger\or \ddagger\or
   \mathsection\or \mathparagraph\or \|\or **\or \dagger\dagger
   \or \ddagger\ddagger \else\@ctrerr\fi}}
\makeatother

\title{Modular Retrieval for Generalization and Interpretation}  % Interpretation

\author{%
  Juhao Liang$^{\includegraphics[width=0.3cm]{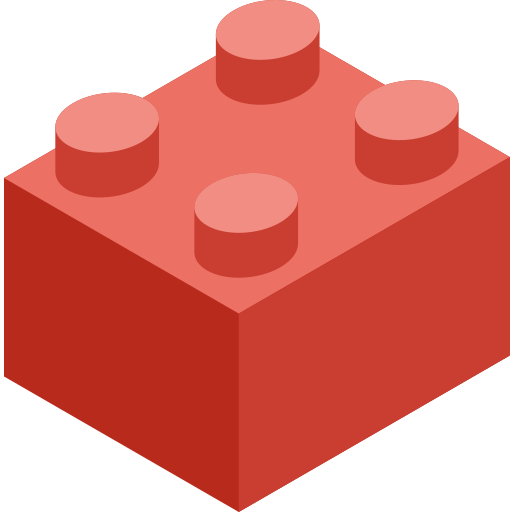}}$,
  Chen Zhang$^{\includegraphics[width=0.3cm]{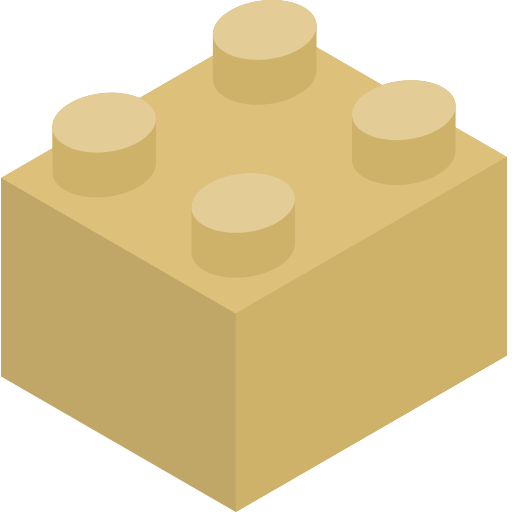}\includegraphics[width=0.3cm]{figures/red_lego.png}}$,
  Zhengyang Tang$^{\includegraphics[width=0.3cm]{figures/red_lego.png}}$,
  Jie Fu$^{\includegraphics[width=0.3cm]{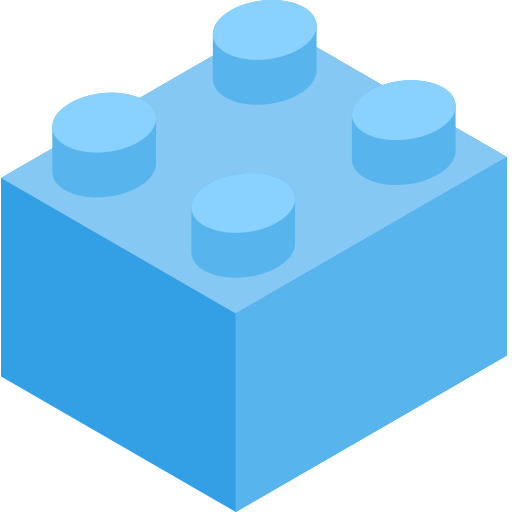}}$, 
  Dawei Song$^{\includegraphics[width=0.3cm]{figures/orange_lego.png}}$, 
  Benyou Wang$^{\includegraphics[width=0.3cm]{figures/red_lego.png}}$\thanks{Benyou is the corresponding author.} \\
   $\includegraphics[width=0.35cm]{figures/red_lego.png}$ SDS \& SRIBD,  The Chinese University of Hong Kong, Shenzhen\\
   $\includegraphics[width=0.35cm]{figures/orange_lego.png}$ Beijing Institute of Technology\\
   $\includegraphics[width=0.35cm]{figures/blue_lego.png}$ Beijing Academy of Artificial Intelligence \\
  \texttt{wangbenyou@cuhk.edu.cn} \\
}

\begin{document}

\maketitle

\begin{abstract}
% key words: information retrieval, modular learning, deep prompt tuning
New retrieval tasks have always been emerging, thus urging the development of new retrieval models. However, instantiating a retrieval model for each new retrieval task is resource-intensive and time-consuming, especially for a retrieval model that employs a large-scale pre-trained language model. To address this issue, we shift to a novel retrieval paradigm called modular retrieval, which aims to solve new retrieval tasks by instead composing multiple existing retrieval modules.
Built upon the paradigm, we propose a retrieval model with modular prompt tuning named REMOP. It constructs retrieval modules subject to task attributes with deep prompt tuning, and yields retrieval models subject to tasks with module composition.
We validate that, REMOP inherently with modularity not only has appealing generalizability and interpretability in preliminary explorations, but also achieves comparable performance to state-of-the-art retrieval models on a zero-shot retrieval benchmark.\footnote{Our code is available at \url{https://github.com/FreedomIntelligence/REMOP}}
  
\end{abstract}

\section{Introduction}

The use of large-scale pre-trained language models (PLMs) has become the de facto standard for various natural language processing (NLP) tasks due to their compelling performance in a wide range of applications, such as sentiment analysis~\cite{MaZS21,ZhangRMW0022,mao2022biases} and named entity recognition~\cite{taher2020beheshti,sun2021rpbert}. Similarly, in retrieval, representative retrieval models~\cite{karpukhin2020dense,xiong2020approximate} also use PLMs as backbones.

However, new retrieval tasks have continuously been emerging due to various requirements, possibly involving distinguished task objectives, domains, and languages. A quick example is personalized retrieval~\cite{dumais2003stuff}, which requires retrieving documents given solely user preferences. These complex tasks bring great challenges for retrieval models. Fortunately, with fine-tuning (FT), retrieval models enhanced by PLMs could be effectively adapted to tasks via tuning all parameters of the model. As a lightweight alternative to FT, parameter-efficient approaches~\cite{houlsby2019parameter} like deep prompt tuning (DPT)~\cite{liu2021pre, li2021prefix} adapt PLMs by only tuning a small set of parameters and have shown performance comparable to FT in retrieval scenarios~\cite{tang2022dptdr}. 
However, such a one-model-to-one-task (one-to-one) retrieval paradigm may fail to handle unseen retrieval tasks due to its limited generalizability, thereby being resource-intensive and time-consuming when the number of tasks explodes.

Inspired by the success of modular learning applied to other NLP tasks~\cite{zhang2019sentiment, pfeiffer2020mad}, in this paper, we formally propose a novel many-modules-to-many-tasks (many-to-many) retrieval paradigm: modular retrieval. Modular retrieval is designed to enhance the generalizability of the retrieval process by combining multiple retrieval modules to solve numerous retrieval tasks. 
Specifically, %instead of picking the related tasks to train a retrieval model for a new task, 
we associate a task into multiple task attributes (e.g., domain) that are exactly in correspondence to reusable modules. 
% we decompose a task into multiple reusable modules that are exactly in correspondence to task attributes (e.g., objective, domain, and language). 
Since we decouple tasks and modules, modular retrieval can be generalized to a new task by selectively composing these modules. In particular, this process does not require any further training.
% Furthermore, the explicit composition of modules improves the interpretability of the retrieval model.

Apart from \textit{generalization}, \textit{interpretation} is another advantage of modular retrieval. By delving deeply into the attributes of each task, we can ground these characteristics to a specific group of parameters, known as a \textit{module}. The composition of these modules can be used in a simple yet effective additive mechanism (see a gradient accumulation analogy for a better understanding).
This process of module decomposition and composition contributes to \textit{decomposability}, a crucial aspect of \textit{transparency} in model \textit{interpretability}, which expects that each sub-component of a network provides an intuitive explanation~\cite{lipton2018mythos}. It is particularly beneficial for various use cases in downstream tasks. For instance, when the need arises to improve certain features, such as updating a domain knowledge module, we can directly locate a grounded module and update it without affecting other parts of the model. More interestingly, we can even add or remove some module parameters in a plug-and-play manner, if necessary.

\begin{figure}[htb]
    \centering
      \includegraphics[width=1.0\textwidth]{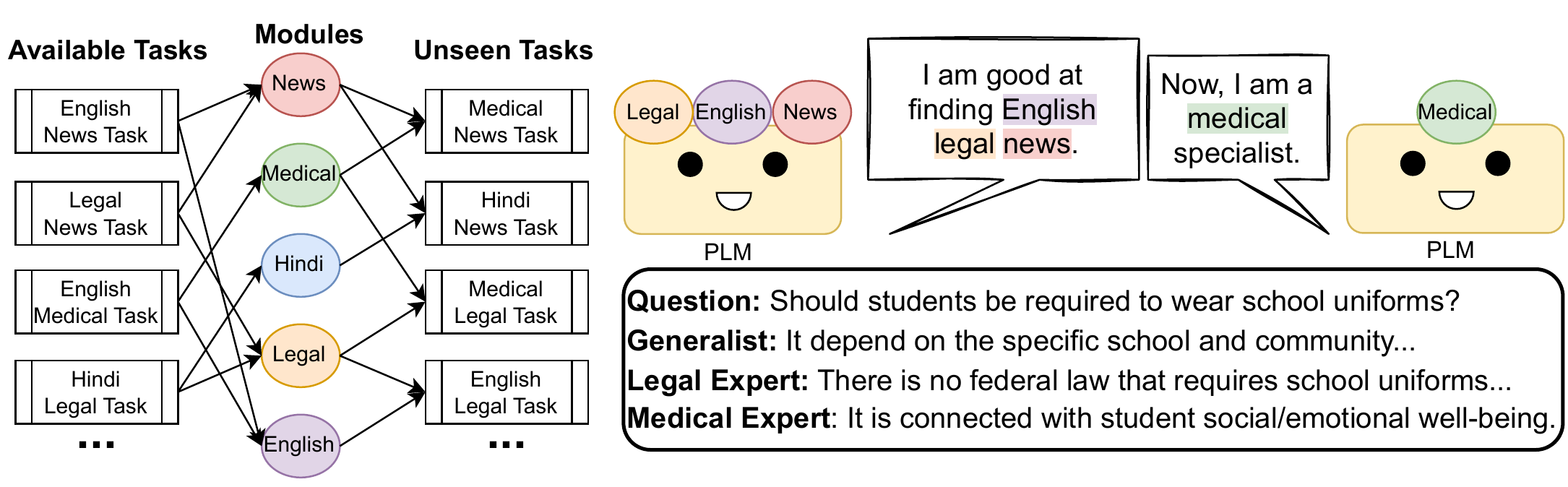}
    \caption{Overview of modular retrieval. Train specific retrieval modules from available tasks and selectively combine them into a PLM to solve complex retrieval tasks, such as personalized retrieval.}
    \label{fig:illustration}
\end{figure}

Based on the paradigm, we propose a new modular retrieval approach, \textbf{REMOP} (\textbf{RE}trieval with \textbf{MO}dular \textbf{P}rompt Tuning), which leverages the DPT technique to construct retrieval modules subject to task attributes and uses the module composition to obtain retrieval models subject to tasks. As shown in Figure \ref{fig:illustration}, REMOP trains several attribute-specific retrieval modules on available tasks and generalizes to unseen tasks via flexible module composition via operating prompts.
% With the plug-and-play feature, the approach only needs simple module operations to deal with complex retrieval scenarios at low cost. 
We conduct multiple explorations to verify the generalizability and interpretability of module composition, and demonstrate its effectiveness on zero-shot retrieval tasks.

To sum up, our contributions are as follows:

\begin{itemize}
  \item We formally introduce a new retrieval paradigm: \textit{modular retrieval} and its merits compared to conventional retrieval.
  \item We implement \textit{modular retrieval} using deep prompt tuning, resulting in a novel modular retrieval approach called \textit{REMOP} that enables a simple-yet-effective module composition. Note that we are the first to leverage modular prompt in general modular training. %facilitate
  \item We explore and validate the generalizability and interpretability of various module compositions in REMOP.
  \item We apply \textit{modular retrieval} in a new application: zero-shot  retrieval. Experiments demonstrated the effectiveness of REMOP.
\end{itemize}

\section{Modular Retrieval}
\label{section: modular retrieval}
\subsection{Definition}

The concept of \textit{modularity}~\cite{fodor1983modularity, ballard1986cortical} has loomed large in the philosophy of psychology since the early 1980s, and it is defined as the correspondence between strongly interconnected components of a system (i.e., modules) and the functions they perform \cite{baldwin2000design, ulrich1995role}. Within the mechanism, each module is tailored for a specific purpose and can be reused consistently. 
There are many studies~\cite{jacobs1991task, rosenbaum2017routing} on the application of modularity in neural networks, and they exploit different approaches to instantiate modules and pursue the design of modular neural networks with \textit{reusability}, \textit{generalizability}, and \textit{interpretability}.

To the best of our knowledge, we are the first to formally define \textit{modular retrieval}, to bring the merits of modularity to retrieval.
\textit{Modular retrieval} is defined as a new information retrieval paradigm that aims to perform retrieval tasks by composing retrieval modules. Each retrieval module is specialized for an attribute of the retrieval task, such as the retrieval goal, the language, or the domain involved. 

The whole modular retrieval process is shown in Algorithm \ref{algorithm: modular retrieval}. Modular retrieval first decomposes a given task into a set of attributes, where each attribute is addressed by a module; and then selects the corresponding retrieval modules according to the decomposed attributes; afterwards, combines the selected retrieval modules with a PLM to form a task-specific model, and uses it to perform the concerned retrieval task.

\begin{algorithm}[t]
    % \renewcommand{\thealgorithm}{}
    % \floatname{algorithm}{}
    \caption{Modular Retrieval }
    \label{algorithm: modular retrieval}
    \begin{algorithmic}
        \STATE \textbf{Given} a retrieval task $t$ and a set of retrieval modules $M$, where module $m_{\alpha}  \in M$ is specialized for retrieval task  attribute $\alpha$. 
        \STATE \textbf{Modular Retrieval Steps:}
        \STATE 1. Decompose the task into a set of modules on task attributes: $t \to A = \{\alpha, \beta, \dots\}$;
        \STATE 2. Select retrieval modules corresponding to the attributes: $M_\textrm{\scriptsize selected} = \left \{ m_{k} \in M | k \in A  \right \} $;
        \STATE 3. Obtain a task-tailored retrieval model by module composition: Retriever$(PLM, M_\textrm{\scriptsize selected})$;
    \end{algorithmic}
\end{algorithm}

\subsection{Module}

Before delving into how tasks can be decomposed into attributes and naturally how corresponding modules can form models for tasks, without losing any generalizability, we formally define the module and its arithmetic thereof.

\subsubsection{Atomic and Compound Modules}

% \subsection{Module Types}
We first define two types of retrieval modules for modular retrieval according to their functionalities: 

\begin{definition}
    % $\textsc{Atom}(\alpha)$
    \textbf{Atomic Module} $\textsc{Module}(\alpha)$ refers to a retrieval module that is specific for one and only one task attribute $\alpha$. For example, a biomedical atomic module is only for solving biomedical-related retrieval tasks, and performs poorly in other domains. 
\end{definition}

\begin{definition}
    \textbf{Compound Module} $\textsc{Module}(\alpha, \beta, \dots\, \theta)$ refers to a retrieval module that is capable of many independent task attributes $\alpha, \beta, \dots, \theta$. For example, using a single science-medical compound module can effectively solve both science-related and medical-related retrieval tasks.
\end{definition}

\subsubsection{Module Arithmetic}

\begin{table}[htp]
\centering
\begin{tabular}{lll}
\toprule
\textbf{Module Arithmetic}
\\ \midrule

Scaling
& $w \times$Module($\alpha$) = Module($w \alpha$)                                           
\\

Addition                      
& Module($\alpha$) + Module($\beta$) = Module($\alpha,\beta)$
\\

Subtraction
& Module($\alpha$,   $\beta$) - Module($\alpha$) = Module($\beta$)                            
\\
\bottomrule
\end{tabular}
% \end{adjustbox}
\caption{Basic module operations for flexible module composition.}
\label{table: basic module operations}
\end{table}

Given a set of retrieval modules, how to enable them with the ability to solve new tasks is an important problem. Although modular retrieval is flexible in this case, we should define three basic modular operations: module scaling, module addition, and module subtraction to realize flexibility, as shown in Table \ref{table: basic module operations}. 

\begin{definition}
\label{definition: module scaling}
\textbf{Module Scaling} is to add a scale weight to a module to control its effectiveness. It is based on the recognition that task attributes have different importance for different tasks. For example, questions asked by a senior scholar can be considered more in-depth than those asked by a beginner.

\end{definition}

\begin{definition}
\label{definition: module addition}
\textbf{Module Addition}, like math addition, is the process of putting together two different (atomic/compound) modules to produce a new (compound) module; and this is for complex retrieval scenarios with multiple task attributes.
\end{definition}

\begin{definition}
\label{definition: module subtraction}
\textbf{Module Subtraction}, as opposed to module addition, is used to remove some functions from a module to obtain a new module that better fits the task. 
For example, when we only have a computer-biomedical compound module and a computer atomic module, we can obtain a pure biomedical atomic module through the subtraction of modules.
\end{definition}

\subsection{Task Decomposition and Module Composition}
\label{section: task decomposition}

\begin{figure}[htp]
    \centering
    \includegraphics[width=0.8\textwidth]{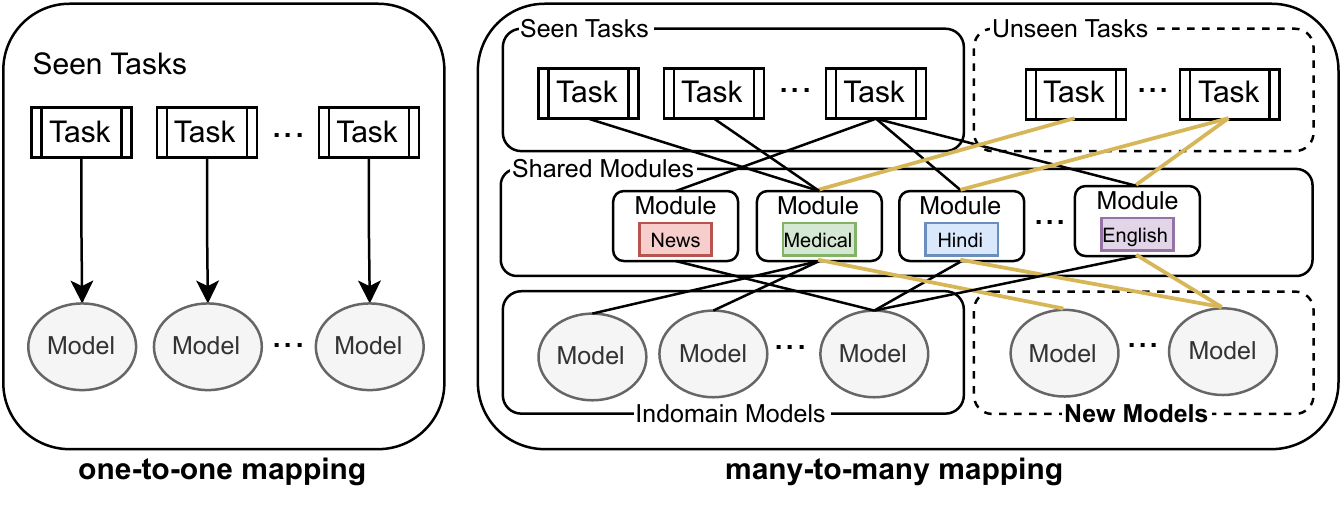}
    \caption{Generalize knowledge from seen retrieval tasks to unseen retrieval tasks by \textit{task decomposition} and \textit{module composition}, thereby improving multi-tasks learning's \textbf{generalizability}.}
    \label{fig:task decomposition}
\end{figure}

One of the advantages of multi-scenario learning \cite{maillard2021multi,asai2022task} is that it could learn the commonness between tasks and therefore easily generalizes to one of the specific tasks. However, it cannot generalize to a new task that was never seen. To overcome this limitation, we propose to decompose a task into multiple attributes, each attribute is associated with a module, as shown in Figure \ref{fig:task decomposition}. It, therefore, results in a many-to-many mapping from modules to tasks. For example, as shown in Table \ref{table: module generalize example}, task $A$ is decomposed into modules $\alpha$ and $\beta$ while task $B$ is decomposed into modules $\alpha$ and $\theta$. After jointly training tasks $A$ and $B$, we have well-trained modules $\alpha$, $\beta$, and $\theta$, by using which, we could infer on many new tasks, e.g., a new task $C$ with $\beta$ and $\theta$ and a new task $D$ with $\alpha$, $\beta$ and $\theta$. Note that $C$ and $D$ are different from $A$ and $B$, but they are composed of an identical module set.
\begin{table}[h]
    \centering
    \begin{tabular}{llll}
    \toprule
    Paradigm & Training Task  & Inference Task \\
    \midrule
    multi-task retrieval & $A$, $B$ & $A$, $B$\\
    modular retreival &  $A$($\alpha$, $\beta$), $B$($\alpha$, $\theta$) & $A$,$B$, {\color{red}$C$($\beta$ $\theta$),  $D$($\alpha$, $\beta$, $\theta$), etc.}\\
    \bottomrule
    \end{tabular}
    \caption{Modular retriever could generalize to {\color{red}{new}} (composed) tasks.}
    \label{table: module generalize example}
\end{table}

An extreme case of modular retrieval will degrade into a multi-task retrieval when module-task mapping degrades to a bijective one, in which case it could not generalize to new tasks.
We could define a bipartite graph between modules and tasks, where one set of vertexes is for \textit{modules} and another is for \textit{tasks}. To balance training efficiency and generalization, each node in both sets in the ideal mapping should not be sparse, in other words, each task should be associated with multiple modules and each module should be associated with multiple tasks. In doing so, not only the commonness between multiple tasks 
(as multi-task learning does) can be learned, but also the generalization of modules can be assured. One could design a human-written mapping based on some heuristics. Alternatively, Section~\ref{section: instruction decomposition} introduces an automatic way for task decomposition.

\section{REMOP: Retrieval with Modular Prompt Tuning}
\label{section: retrieval using deep modular prompts}

\subsection{Basic Architecture}

\begin{figure}[htp]
    \centering
    \includegraphics[width=1.0\textwidth]{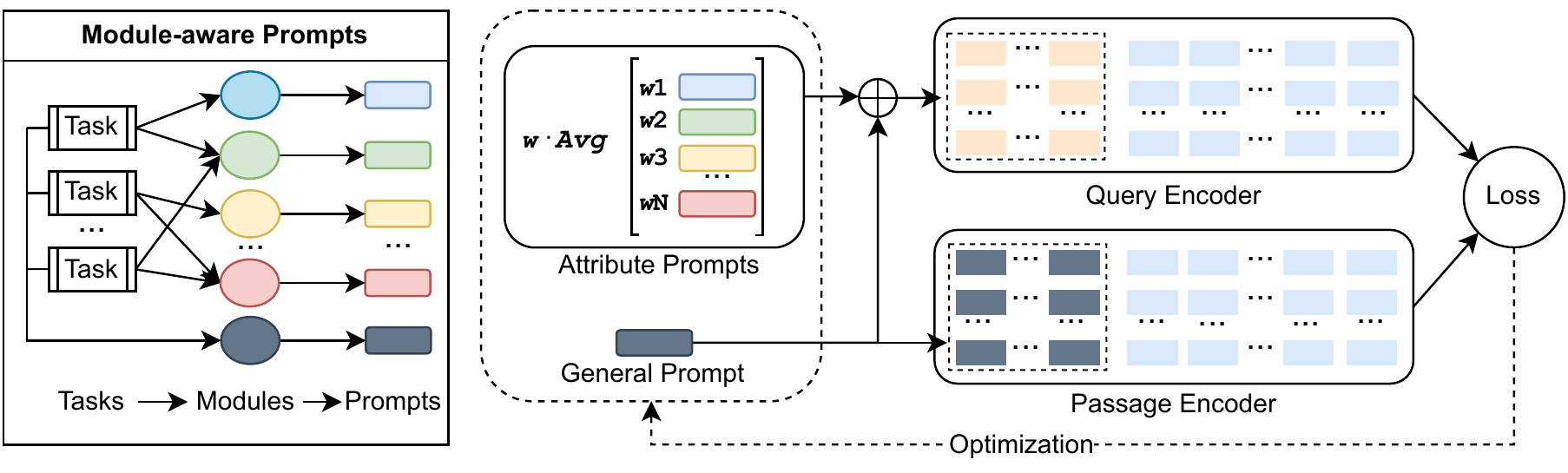}
    \caption{Overview of REMOP framework, built upon a dual-encoder framework. Blue blocks represent the frozen PLM backbone, dark gray blocks represent general prompts, and other colored blocks represent different attribute prompts. }
    \label{fig:framework of REMOP}
\end{figure}

\paragraph{Problem Definition}
In our settings, modular retrieval aims to retrieve related passages according to a given query and its task attributes.\footnote{Sometimes, a task is described by a textual instruction.} It requires the learning of attribute knowledge into corresponding modules and the combination of selected modules to perform a new retrieval task.

\paragraph{Architecture} 
Based on the proposed retrieval paradigm, we present an architecture shown in Figure \ref{fig:framework of REMOP}.
It is built upon a dual-encoder framework, which is comprised of a query encoder and a passage encoder. The function of the model is to retrieve the passages with encoded vectors closest to the vector of the given query. Specifically, we use a PLM-based dual-encoder to separately encode the query and the passages and computes their vector similarities\cite{mussmann2016learning, ram2012maximum} in the dense vector space, the similarity score is computed as the inner product as previous works indicated \cite{karpukhin2020dense}:
\begin{equation}
   s(q, p)=E_{q}(q) \cdot E_{p}(p) 
\end{equation}

\paragraph{Prompt Tuning Implementation}
Unlike conventional dense passage retrieval models, REMOP adds deep modular prompts to the two encoders respectively to steer the retrieval direction specifically. The deep modular prompt essentially is a prefix prompt \cite{liu2021pre,li2021prefix}, but in the modular retrieval scenario, it is also a container of task attribute knowledge and the realization of the retrieval module. The selection of a deep prompt will be conditioned on a task and its associated attributes. In general, there are many parameter-efficient ways to implement attribute-specific modules, such as deep prompts \cite{liu2021pre,li2021prefix} (used in this work), adapters \cite{rebuffi2018efficient,pfeiffer2020mad} (previously used in multi-tasking learning and also nowadays in fine-tuning), experts in Mixture-of-Expert architecture \cite{artetxe2021efficient} or LoRA \cite{hu2021lora}. The main rationale for using deep prompt is that we believe that attribute-specific information should be stored in self-attention networks \cite{li2021prefix} while deep prompt could directly intervene in the computation of the self-attention network.  

% As a comparison, Feed-forward network usually store domain knowledge (e.g. biology, medicine, or math) that are mostly orthogonal to task specificity, see related work in \cite{geva2020transformer,dai2021knowledge,meng2022locating}.

\subsection{Separating Specificity from Commonness}

We divide deep modular prompts into two types: a \textit{general prompt} and many \textit{attribute prompts}. The former learns the commonness among tasks while the latter learns the specificity of each attribute.  

\textbf{General Prompt} is used to store the general knowledge of the retrieval tasks, and can also be considered as an adapter between attribute prompts and the PLM, because all attribute prompts are not directly attached to the PLM, but attached to the general prompt.

\textbf{Attribute Prompt} is used to store specific task attribute knowledge. It has the ability of module operations mentioned above, and can be flexibly combined with each other as requirements. While both the query encoder and passage encoder come with a general prompt, attribute prompts are only allowed to attach on the query encoder. This is because otherwise we need to stores many encoded passage vectors for tasks with different attributes. We can actually reuse the same encoded passages, avoiding repeat computation and reducing the cost of storage. 

\textbf{Necessity of the General and Attribute Prompts}
There are three types of parameters: the frozen \textit{PLM backbone}, the trainable \textit{general prompt} and trainable \textit{attribute prompts}.
Since the PLM backbone is frozen to align standardized backbones among (classification or retrieval) tasks, the only trainable parameters sharing between tasks are general prompt, without which REMOP cannot learn the commonness between tasks. Attribute prompts are used to learn the task specificity, or more accurately, attribute specificity.

% \chen{3.3 and 3.4 have a certain degree of overlap with 2.3, consider using instruction decomposition and prompt composition or something alike to denote that they are at implementation level now.}
\subsection{Instruction Decomposition}
\label{section: instruction decomposition}

Inspired by instruction tuning \cite{wei2021finetuned,sanh2021multitask,ouyang2022training,chung2022scaling,wang2022super}, we use task instructions to build a task module mapping. Concretely, we decompose  task instructions into multiple task attribute labels, as shown in Table \ref{table: examples of task attributes extraction.}, and by leveraging the task attributes labels, we can consider that a task consists of multiple modules. Consequently, we transform a retrieval with instruction task \cite{asai2022task} into a modular learning task \cite{pfeiffer2023modular}, where a module is refer to one or multiple task attributes. To sum up, we utilize the modular learning method to solve the retrieval with instruction tasks in order to obtain better interpretability and flexibility. 

\begin{table}[htp]
\centering
\begin{tabular}{ll}
\hline
\textbf{Task} & \textbf{Instruction}                                                          \\ \hline
NF Corpus        & Retrieve \colorbox{cblue}{scientific paper} paragraph to \colorbox{cgreen}{answer this question.}                   \\
Arguana          & Retrieve an argument that \colorbox{cyellow}{counter argues} the following paragraph.              \\
SciFact          & Retrieve a \colorbox{cblue}{scientific paper} sentence to \colorbox{cyellow}{verify if the following claim is true.} \\
DBPedia          & Retrieve an \colorbox{cred}{Wikipedia} introduction paragraph of the following entity.          \\ \hline
\textbf{Attribute}      & \colorbox{cblue}{Science}, \colorbox{cgreen}{QA}, \colorbox{cyellow}{Fact-Checking}, \colorbox{cred}{Wikipedia}                                         \\ \hline
\end{tabular}
\caption{Examples of instruction decomposition for retrieval tasks: NF Corpus~\cite{boteva2016full}, Arguana~\cite{wachsmuth-etal-2018-retrieval}, SciFact~\cite{wadden2020fact}, and DBPedia~\cite{hasibi2017dbpedia}.}
\label{table: examples of task attributes extraction.}
\end{table}

\subsection{Modular Prompt Composition}
\label{sec:attribute combination}
% Distributed Gradient 

It is a crucial problem about how to implement the module operations mentioned in the modular retrieval paradigm in REMOP. Weighted averaging \cite{mcmahan2016federated,mcmahan2017communication} is a commonly used approach to combine different models in federated learning \cite{li2020federated,konevcny2016federated}. More specifically, weighted averaging involves assigning each local model a weight that reflects its performance on a validation set or some other metric, and then combining the models by taking a weighted average of their parameters. In the modular retrieval case, we can consider a pre-trained language model with different attribute prompts as several independent models, and using the weighted averaging method to combine them can be considered as building a new retrieval model with all the source models' ability. To this end, we apply the weighted averaging method to the trained attribute prompts for module compositions.  

For instance $x$, the final prefix prompt is $P_\textrm{final}(x)$, which is calculated by a shared general prompt $P_\textrm{general}$ and $N$ attribute prompts $P_{1} \dots P_{N}$, where $N$ indicates the total number of the kinds of attributes in the whole training datasets. Besides, the instance $x$ has $N$ weighting values $w_{x1} \dots w_{xN}$ corresponding to the task attributes. Therefore, the final prompt $P_\textrm{final}$ for the instance $x$ is calculated as following:
\begin{equation}
    P_\textrm{final}(x) = P_\textrm{general} + \frac{w}{N} \sum_{i}^{N}w_{xi}P_{i}
\label{formula: prompt composition}
\end{equation}

where $w$ indicates the scale weight of the synthesized attribute prompt. Section.~\ref{subsection: module addition} experimentally demonstrates the effectiveness of module scaling, module addition, and module subtraction by using such a weighted averaging composition approach. 
% By extending addition, one can implement other module arithmetic operations: module scaling and module subtraction.

% \paragraph{An Intuitive Explanation:}   
% \benyou{accumulated gradient weight vs learning rate? weight =1  explain}
% ‘delta-tuning’, where ‘delta’ a mathematical notation often used to denote changes, is borrowed to refer to the portion of parameters that are ‘changed’ during training.

\paragraph{An Intuitive Explanation:}  
Eq.~\ref{formula: prompt composition} inherently assumes that\textit{ modular prompts are \textbf{additive}}. We argue that the rationale behind \textit{additive} prompts is nontrivial.
In a parameter-efficient fashion, prompts are the only trainable parameters while the basic backbone is frozen.  From an intuitive perspective, learned specific prompts w.r.t a specific module could be considered as specific \textit{accumulated gradients on the general prompt}  performed on all related tasks w.r.t the  module. In other words, specific prompts, previously considered as network parameters, could, at least in some senses, be treated as gradients of general prompts; note that they are the same shape. Gradient accumulation is a way to train deep learning models that involves adding up gradients over multiple mini-batches before updating the model's weights; such gradients are \textbf{additive}. Therefore, weight average between multiple prompts could be seen as accumulating `related gradients learned from related tasks' while excluding unrelated ones, namely, selecting approximate modular prompts during inference.

% , although one task might be associated with multiple prompts,

% An \textit{intuitive understanding} of the prompt composition method (i.e., weighted averaging) is gradient accumulation over multiple training for different tasks. 
% Gradient accumulation is a way to train deep learning models that involves adding up gradients over multiple mini-batches before updating the model's weights. 
For prompt training in REMOP, we use data with the same task attributes to train an attribute prompt and directly add it to the general prompt, so that the former can be considered as an update of the latter one (i.e., delta value). Then, we add a weight to the \textit{attribute prompt}, similar to imposing a `learning rate' upon the \textit{the gradients}, to control the effect of the update in this training batch. The higher the weight, the faster the model is updated, which may lead to task overfitting, or even move away from the optimal point. In contrast, a lower scale weight may steer the model more general than specific. Therefore, finding an appropriate scaling weight is crucial to improve the generalization and adaptability of the retrieval model.
% And for prompt composition, we can think of each task as a training batch, accumulating their parameter updates together to form a whole training.

% Given a pre-trained model and several downstream tasks, a conventional method is to use the gradient descent algorithm to update the parameters of the given model over multiple training iterations for each task. The parameter update increment for each task can be considered task-specific knowledge. From this point of view, accumulating the updates of several tasks is like training the original model over multiple tasks sequentially and learning knowledge from several tasks.

% Gradient accumulation is a technique used in machine learning where gradients computed during forward and backward passes on a mini-batch of data are accumulated across multiple iterations, instead of updating the model's parameters after every iteration.

% \subsection{Model Pre-training}
\subsection{Modular Prompt Training}
% \chen{seems that textbf is misused for paragraph here and somewhere else i doubt.}

% \paragraph{Mixed-Task Training}
% two-stage training: general first and attriubte later
% two-speed learning rate
The proposed method needs to train both a general prompt and attribute prompts on multiple source retrieval tasks. We divide the training into two phases: general prompt training and attribute prompt training. The first stage uses general retrieval tasks (e.g., MS-MARCO~\cite{nguyen2016ms}) to solely train the general prompt, which is to learn commonsense knowledge and guarantee its basic general retrieval ability. 
In the second phase, we jointly train the attribute prompts and the general prompts on all the available tasks. Specifically, we employ a lower learning rate for the trained general prompt and a higher learning rate for the attribute prompts, which is objective to learn specific knowledge into the attribute prompts and fine-tune the general prompt to better bridge the attribute prompt with PLM.
% In the second phase, we independently train each retrieval module on their corresponding tasks. For example, we select the tasks related to Wikipedia content to train a Wikipedia-specific retrieval module. The two-stage training strategy makes module training more independent and provides feasibility for distributed parallel training while maintaining compatibility among modules.

\section{Exploration on Modularity}
\label{section: module operation exploration}
In this section, we separately verify the feasibility of module arithmetic of deep modular prompts implemented by REMOP through conducting three exploratory experiments. After that, there is a case study to show how to leverage these three module operations to flexibly compose different modules to solve tasks in complex scenarios. The overview of the experiment design is shown in Table \ref{table: module operation experiments}.

\begin{table}[htp]
\begin{adjustbox}{width=1.0\columnwidth,center}
\begin{tabular}{lllll}
\toprule
& \textbf{Training} & \textbf{Module Operations} & \textbf{Evaluation}              \\ \midrule
\textbf{Scaling}
& Module($\alpha$)                                                                    & \begin{tabular}[c]{@{}c@{}}$w$Module($\alpha$)\end{tabular}   

& Data\{$\alpha$\}                                              
\\ \hline

\textbf{Addition}               
& \begin{tabular}[c]{@{}c@{}}
Module($\alpha$)  \\      
Module($\beta$)
\end{tabular}                
& Module($\alpha$) + Module($\beta$)              
& Data\{$\alpha$, $\beta$\}                                                         \\ \hline

\textbf{Subtraction}                  
& \begin{tabular}[c]{@{}l@{}}
Module($\alpha$, $\beta$)  \\
Module($\alpha$)
\end{tabular}
& Module($\alpha$,   $\beta$) - Module($\alpha$)                                
& Data\{$\beta$\}                                                                \\ \hline

\textbf{Combination} & \begin{tabular}[c]{@{}c@{}}Module($\alpha$)\\      Module($\beta$)\end{tabular} &
\begin{tabular}[c]{@{}c@{}}$w_{1}$Module($\alpha$) + $w_{2}$Module($\beta$)\end{tabular} 
& \begin{tabular}[c]{@{}c@{}}$w_{1}$Data\{$\alpha$\}+$w_{2}$ Data\{$\beta$\}\end{tabular} \\
\bottomrule
\end{tabular}
\end{adjustbox}
\caption{Overview of module arithmetic experiments.}
\label{table: module operation experiments}
\end{table}

\subsection{Experimental Setup}
\label{section: exploration experiments setup}

We use a RoBERTa-large model\cite{liu2019roberta} as the backbone PLM for the following experiments due to its outstanding performance compared with other publicly available large language models. Besides, a previous research\cite{tang2022dptdr} has shown that directly applying deep prompt tuning in dense passage retrieval largely underperforms fine-tuning methods, and it is helpful and necessary to do the retrieval-oriented intermediate pretraining (RIP)\cite{tang2022dptdr} for the large language model. Therefore, we use the RoBERTa-large model which has been pre-trained on MS-MARCO\cite{nguyen2016ms} by RIP as our final fixed backbone model. The length of all prefix-prompts is set to 128, and the training epoch is set to 5. 

For each experiment in this section, we use \textbf{100k data for training} and \textbf{1k data for evaluation}, and all data are sourced from BERRI \cite{asai2022task} and LOTTE \cite{santhanam2021colbertv2}. Following the instruction decomposition method described in Section~\ref{section: instruction decomposition}, we extract the task attributes from the instructions for each task and map them to the corresponding modules. The results reported are based on the performance metric of Normalized Discounted Cumulative Gain at 10 (i.e., NDCG@10).

\subsection{Module Scaling}
\label{subsection: module scaling}
As defined in Definition \ref{definition: module scaling}, module scaling is to add a scale weight to control the effect of an attribute module on the generalist model. If the weight is set too low (close to 0), the model will be more general instead of task-attribute-specific, while if it is set too high (close to 1), the model tends to overfit the training data, leading to poor generalizability. In order to find a proper scale weight, we test the effect of seven different scale weights on five downstream tasks which have different task attributes. As shown in Table \ref{table: result of module scaling}, the results show that using module weights close to 0 or close to 1 will result in poor zero-shot retrieval ability. Also, 0.5 is a module weight that performs well in most experiments. Therefore, we select 0.5 as the default module scale weight $w$ for the final attribute prompt, and it is applied in the remaining experiments.

\begin{table}[htp]
\centering
\begin{adjustbox}{width=\columnwidth,center}
\begin{tabular}{lccccccc}
\toprule
\textbf{Task / Module} & General   & 0.1×$M_{\alpha}$   & 0.3×$M_{\alpha}$   & 0.5×$M_{\alpha}$   & 0.7×$M_{\alpha}$   & 0.9×$M_{\alpha}$   & 1.0×$M_{\alpha}$   \\ \midrule
\colorbox{cgreen}{QA (1k)}           & 0.315 & 0.325 & 0.342 & \textbf{0.357} & 0.350 & 0.300 & 0.267 \\
\colorbox{cblue}{Science (1k)}          & 0.152 & 0.154 & 0.155 & \textbf{0.155} & 0.153 & 0.147 & 0.144 \\
\colorbox{corange}{Summarization (1k)}          & 0.152 & 0.153 & \textbf{0.153} & 0.151 & 0.147 & 0.136 & 0.128 \\
\colorbox{cred}{Wikipedia (1k)}         & 0.448 & 0.460 & 0.474 & \textbf{0.474} & 0.451 & 0.379 & 0.312 \\
\colorbox{cyellow}{Fact-Checking (1k)}           & 0.134 & 0.152 & 0.181 & \textbf{0.194} & 0.186 & 0.164 & 0.145 \\ \bottomrule
\end{tabular}
\end{adjustbox}
\caption{Results of module scaling experiments. Using a corresponding attribute module $M_{\alpha}$ for each task: \colorbox{cgreen}{Module(QA)}, \colorbox{cblue}{Module(Science)}, \colorbox{corange}{Module(Summarization)}, \colorbox{cred}{Module(Wikipedia)}, \colorbox{cyellow}{Module(Fact-Checking)}, compare the performance of different module scale weights.}
\label{table: result of module scaling}
\end{table}

\subsection{Module Addition}
\label{subsection: module addition}
To find the best way to implement the module addition arithmetic described in Definition \ref{definition: module addition}, which is to combine two modules into a new module, we try two simple aggregation operations: sum aggregation (Sum) and average aggregation (Avg). 
Specifically, we train two retrieval modules separately, a Wikipedia module (wiki) and a Science module (sci), and evaluate the performance of models with different module settings in a corresponding downstream task that is associated with both science and Wikipedia. Table \ref{table: module addition experiment} shows that the effect of using one of the trained modules is superior to the generalist model, whereas the performance of combining modules through sum aggregation is the worst. The module synthesized using average aggregation has the best performance among the experiments. To this end, the average aggregation is chosen as the implementation of module addition in REMOP.

\begin{table}[htp]
\centering
\begin{adjustbox}{width=\columnwidth,center}
\begin{tabular}{cccccc}
\toprule
\textbf{Task / Module} & \colorbox{lightgray}{General} & \colorbox{cblue}{Module(sci)} & \colorbox{cred}{Module(wiki)} & Sum(\colorbox{cblue}{sci}, \colorbox{cred}{wiki}) & Avg(\colorbox{cblue}{sci}, \colorbox{cred}{wiki}) \\ \midrule
\colorbox{cblue}{Science}+\colorbox{cred}{Wikipedia} (1k)         & 0.357         & 0.382        & 0.376         & 0.353                   & \textbf{0.385}              \\ \bottomrule
\end{tabular}
\end{adjustbox}
\caption{Results of module addition experiments.}
\label{table: module addition experiment}
\end{table}

\subsection{Module Subtraction}
\label{subsection: module subtraction}
As defined in Definition \ref{definition: module subtraction}, module subtraction aims at creating a new module by removing useless knowledge from a given module. Following Table \ref{table: module operation experiments}, we train two modules: a Wikipedia-Fact-Checking compound module and a Wikipedia atomic module, namely \textit{Module(wiki, fc)} and \textit{Module(wiki)} respectively. Given an unseen Fact-Checking retrieval task, we compare and analyze the effects of different combinations of the above two trained modules. As shown in Table \ref{table: module subtraction experiment}, results show that even though the synthetic module is not better than the compound module, it still outperforms the generalist model and the atomic module, demonstrating the effectiveness of the module subtraction arithmetic.

\begin{table}[htp]
\centering
\begin{adjustbox}{width=\columnwidth,center}
\begin{tabular}{ccccc}
\toprule
\textbf{Task / Module} & \colorbox{lightgray}{General} & Module(\colorbox{cred}{wiki}) & Module(\colorbox{cred}{wiki}, \colorbox{cyellow}{fc}) & Module(\colorbox{cred}{wiki}, \colorbox{cyellow}{fc})-Module(\colorbox{cred}{wiki}) \\ \midrule
\colorbox{cyellow}{Fact-Checking (1k)}        & 0.131         & 0.172        & \textbf{0.195}                   & 0.187              \\ \bottomrule
\end{tabular}
\end{adjustbox}
\caption{Results of module subtraction experiments.}
\label{table: module subtraction experiment}

\end{table}

\subsection{Module Arithmetic Combination}
\label{subsection: module combination}
Based on previous experiments, we observed that not all task attributes are equally important for the same task. For example, in module addition experiments (Section~\ref{subsection: module addition}), the science atomic module has better performance than the Wikipedia atomic module for the given Science-Wikipedia retrieval task. 
To find out the primary reason behind the difference, in this part, we investigate the relationship between module combination weights and task attribute proportions.
First, we use a QA-specific task and a Wikipedia-specific task to synthesize new QA-Wikipedia tasks with different synthesis ratios. Then, leveraging a trained QA module $M_\textrm{\scriptsize QA}$ and a trained Fact-Checking module $M_\textrm{\scriptsize FC}$, we solve the synthesis tasks under different module combination settings. 

As shown in Table \ref{table: module flexible combination results}, combining two modules with different scale weights can have different performance for the same task, and the ratio of module combinations to task combinations can roughly be considered to be linearly related. It shows that it is reasonable to assign different scale weights for task attributes to obtain a more tailored model for a given task. 

\begin{table}[htp]
\centering
\begin{adjustbox}{width=\columnwidth,center}
\begin{tabular}{c|cccccc}
\toprule
\textbf{Task / Module} & \colorbox{lightgray}{General} & \colorbox{cgreen}{$M_\textrm{\scriptsize QA}$} & \colorbox{cgreen}{0.7$M_\textrm{\scriptsize QA}$}+\colorbox{cyellow}{0.3$M_\textrm{\scriptsize FC}$} & \colorbox{cgreen}{0.5$M_\textrm{\scriptsize QA}$}+\colorbox{cyellow}{0.5$M_\textrm{\scriptsize FC}$} & \colorbox{cgreen}{0.3$M_\textrm{\scriptsize QA}$}+\colorbox{cyellow}{0.7$M_\textrm{\scriptsize FC}$} & \colorbox{cyellow}{$M_\textrm{\scriptsize FC}$} \\ 
\midrule
\colorbox{cgreen}{0.7QA}+\colorbox{cyellow}{0.3FC} (1k)        & 0.1827            & 0.1960           & 0.2034                    & 0.2061                    & \textbf{0.2072}                    & 0.2040           \\ %\hline
\colorbox{cgreen}{0.5QA}+\colorbox{cyellow}{0.5FC} (1k)        & 0.1406            & 0.1570           & 0.1616                    & \textbf{0.1633}                    & 0.1633                    & 0.1600           \\ %\hline
\colorbox{cgreen}{0.3QA}+\colorbox{cyellow}{0.7FC} (1k)        & 0.1003            & 0.1214           & 0.1242                    & 0.1247                    & \textbf{0.1252}                    & 0.1220           \\ \bottomrule
\end{tabular}
\end{adjustbox}
\caption{Results of module arithmetic combination experiment.}
\label{table: module flexible combination results}
\end{table}

\section{Generalizing to New Tasks: Zero-shot Retrieval}
\label{section: zero-shot retrieval application}
In this section, we evaluate the generalizability of REMOP in zero-shot retrieval applications. Specifically, we train retrieval modules on a dataset collection containing task instructions written by experts, and test the performance on a zero-shot retrieval benchmark. The experimental settings are shown on Section~\ref{section: zero-shot settings} and the results are demonstrated in Section~\ref{section: zero-shot results}. Section~\ref{section: zero-shot analysis} analyzes the experimental results and gives a conclusion on this experiment.

\begin{table}[htp]
\centering
\begin{adjustbox}{width=1.0\columnwidth,center}
\renewcommand{\arraystretch}{1.2}
\begin{tabular}{ll|ll}

\toprule
\multicolumn{4}{c}{\textbf{Retrieval Modules on Training Phase (from BERRI)}} \\ \midrule
\textbf{Module} & \multicolumn{1}{l|}{\textbf{\#q in train}} & \textbf{Module} & \textbf{\#q in train} \\ \hline
\colorbox{cgreen}{QA}              & 2,355,072    & Dialogue    & 159,064             \\ \hline
\colorbox{corange}{Summarization}   & 921,266      & \colorbox{cyellow}{Fact-Checking} & 140,438           \\ \hline
\colorbox{cred}{Wikipedia}       & 831,003      & \colorbox{cblue}{Science}      & 123,544            \\ \hline
News            & 756,275      & Caption      & 70,643             \\ \hline
Medical         & 324,520      & Legal      & 2,256                \\ \hline
Sentence-Paraphrase & 185,859  & \colorbox{lightgray}{General} & 1,415,211 \\ 
\toprule
\multicolumn{4}{c}{\textbf{Retrieval Tasks on Evaluation Phase (from BEIR)}} \\ \midrule
\textbf{Task} & \multicolumn{1}{l|}{\textbf{Attributes (\#q in test)}} & \textbf{Task}   & \textbf{Attributes (\#q in test)} \\ \hline
Arguana (ARG) \cite{wachsmuth-etal-2018-retrieval}          & \colorbox{cyellow}{Fact-Checking} (1,406) & 
NF Corpus (NFC) \cite{boteva2016full}                       & \colorbox{cblue}{Science}+\colorbox{cgreen}{QA} (3,237)            \\ \hline
Climate-Fever (CLI) \cite{diggelmann_climate-fever_2021}    & \colorbox{cyellow}{Fact-Checking} (1,535) & 
Touche-2020 (TOU) \cite{bondarenko2020overview}             & \colorbox{cyellow}{Fact-Checking} (49)                   \\ \hline
DBPedia (DBP) \cite{hasibi2017dbpedia}                      & \colorbox{cred}{Wikipedia} (467) & 
SciDocs (SCD) \cite{cohan-etal-2020-specter}                & \colorbox{cblue}{Science}+\colorbox{corange}{Summarization} (1,000)           \\ \hline
FIQA (FQA) \cite{maia201818}                                & \colorbox{cgreen}{QA} (3,452) & 
SciFact (SCF) \cite{wadden2020fact}                         & \colorbox{cblue}{Science}+\colorbox{cyellow}{Fact-Checking} (1,109)             \\ \hline
TREC-COVID (TREC) \cite{voorhees2021trec}                   & \colorbox{cblue}{Science}+\colorbox{cgreen}{QA} (50)  & 
                  &   \\ 
\bottomrule
\end{tabular}
\renewcommand{\arraystretch}{1}
\end{adjustbox}
\caption{Data statistics in zero-shot retrieval experiments. \# in train/test indicates the number of queries used for module training and task evaluation. Note that not all the trained modules are used for evaluation.}
\label{table: zero-shot experiment data}
\end{table}

% LOTTE-pooled \cite{santhanam2021colbertv2}
% \begin{tabular}[l]{@{}l@{}}Sentence-Paraphrase (1,071)\\QA (3,869)\\Wikipedia (1,585)\\Science (617)\end{tabular}

\subsection{Experimental Settings}
\label{section: zero-shot settings}
\paragraph{Datasets and Metrics} 
% The data statistics in zero-shot retrieval experiments are shown in Table \ref{table: zero-shot experiment data}.
In the training phase, we use BERRI \cite{asai2022task} for module training. BERRI is a large-scale retrieval dataset with expert-written task instructions; and by decomposing the given task instructions, we can obtain a list of task attributes for each task and train the corresponding retrieval modules following the above mentioned two-stage training strategy.
In the evaluation phase, we evaluate the performance of the trained modules on a zero-shot retrieval benchmark: \textbf{BEIR} \cite{thakur2021beir}. 
Following \cite{asai2022task,dai2022promptagator}, we exclude Natural Questions, MS MARCO, HotpotQA, FEVER and CQADupStack from evaluation tasks.
We map the remaining tasks to one or more retrieval modules according to their task instructions, and the data statistics of evaluation is shown at the lower part of Table \ref{table: zero-shot experiment data}. We report the official NDCG@10 metric on BEIR.

% and Success@5 (Recall@5) on LOTTE-pooled .
% \textbf{LOTTE-pooled} \cite{santhanam2021colbertv2}

\paragraph{Model Settings}
% In order to show the robustness of REMOP on different backbones, we implement REMOP in this experiment using two PLMs as backbones respectively: Condenser-base \cite{gao2021unsupervised} and RoBERTa-large (RIP fine-tuned version) \cite{liu2019roberta,tang2022dptdr}. Following the previous works \cite{tang2022dptdr}, prompt length is set to 128, learning rate is set to 7e-3 for both general and attribute module training. And we adopt the instruction-unfollowing negative sampling to select 5 negative passages for each positive document as BERRI did. The rest of hyper-parameter settings is following coCondenser \cite{gao2021unsupervised} (e.g. warm-up ratio, and mixed-precision training). All training and inference are conducted on a single server with 4 GPUs. 

We implement REMOP in this experiment using MS-MARCO\cite{nguyen2016ms} pre-trained condenser-base model \cite{gao2021unsupervised} as the PLM backbone. Following the previous work \cite{tang2022dptdr}, prompt length is set to 128, learning rate is set to 7e-3 for general prompt training in phase 1 and attribute prompt training in phase 2, the learning rate for phase 2 general prompt training is set to 7e-6. And we adopt the instruction-unfollowing negative sampling to select 5 negative passages for each positive document as BERRI did \cite{asai2022task}. The rest of hyper-parameter settings is following coCondenser \cite{gao2021unsupervised} (e.g., warm-up ratio, and mixed-precision training). All training and inference are conducted on a single server with 4 GPUs. 

\paragraph{Baselines}
Essentially, we empirically found that the retrieval performance of REMOP strongly depends on the PLM backbone. For this reason, the main research objective in this part is to find out if REMOP can enhance the original PLM backbone's generalizability on zero-shot retrieval tasks. We choose a pre-trained condenser as the backbone and compare the zero-shot retrieval performance of DPR (fine-tune all parameters), DPR with single prompt (training a single prompt, following DPTDR \cite{tang2022dptdr}) and REMOP (fine-tune multiple specific modular prompt).
Besides, we also demonstrate the results of other widely used retrieval methods for readers' reference, including a lexical retrieval model, sparse retrieval models, and multiple dense retrieval models; these results are reported from BEIR and BERRI \cite{thakur2021beir,asai2022task}.

% Two kinds of backbone are used to make the results more convincing: a coCondenser-base and a RoBERTa-large model. For each backbone, we compare the performance of the fine-tuned version, backbone with single prompt version (DPTDR) and backbone used for REMOP version. Besides, we also show the results of other widely used retrieval methods for readers' reference: BM25, Contriver, UPR, Contriver(MS), ColBERT-v2, GTR-11B, TART-dual. 

\subsection{Experimental Results}
\label{section: zero-shot results}

\begin{table}[htp]
\centering
\begin{adjustbox}{width=\columnwidth,center}
\renewcommand{\arraystretch}{1.2}
\begin{tabular}{llcccccccccc}
\toprule
% \multicolumn{2}{c}{\textbf{Model}}     & \multicolumn{10}{c}{\textbf{BEIR}} \\ \midrule
% \textbf{Method} &  \textbf{PLM} (\# of trainable parameter) &  \textbf{TREC} & \textbf{NFC}  & \textbf{FQA}  & \textbf{ARG}  & \textbf{TOU}  & \textbf{DBP}  & \textbf{SCD}  & \textbf{CLI}  & \textbf{SCF}  & \textbf{Avg.}  \\ 
\textbf{Method} &  \textbf{PLM} (\# Params) &  \textbf{TREC} & \textbf{NFC}  & \textbf{FQA}  & \textbf{ARG}  & \textbf{TOU}  & \textbf{DBP}  & \textbf{SCD}  & \textbf{CLI}  & \textbf{SCF}  & \textbf{Avg.}  \\ 
 &   &  \colorbox{cblue}{SCI}\colorbox{cgreen}{QA} & \colorbox{cblue}{SCI}\colorbox{cgreen}{QA}  & \colorbox{cgreen}{QA}  & \colorbox{cyellow}{FC}  & \colorbox{cyellow}{FC}  & \colorbox{cred}{WIKI}  & \colorbox{cblue}{SCI}\colorbox{corange}{SUM} & \colorbox{cyellow}{FC}  & \colorbox{cblue}{SCI}\colorbox{cyellow}{FC}  & \\
\midrule
BM25                & - & 65.6 & 32.5 & 23.6 & 31.5 & 36.7 & 31.3 & 15.8 & 21.3 & 66.5 & 36.0 \\ \hline
DeepCT              & BERT-base (110M) & 40.6 & 28.3 & 19.1 & 30.9 & 15.6 & 17.7 & 12.4 & 6.6 & 63.0 & 26.2 \\  
SPARTA              & DistilBERT (66M) & 53.8 & 30.1 & 19.8 & 27.9 & 17.5 & 31.4 & 12.6 & 8.2 & 58.2 & 28.8 \\  
% docT5query          & T5-base (220M) & 71.3 & 32.8 & 29.1 & 34.9 & 34.7 & 33.1 & 16.2 & 20.1 & 67.5 & 37.7 \\  
\hline

Contriever          & BERT-base (110M) & 27.4 & 31.7 & 24.5 & 37.9 & 19.3 & 29.2 & 14.9 & 15.5 & 64.9 & 29.3           \\
% UPR                 & T0-3B (3B) & 60.4 & 33.3 & 45.0 & 50.3 & 21.3 & 33.8 & 17.3 & 9.5  & 69.6 & 37.8             \\ 
% Contriever (MS)     & BERT-base (110M) & 59.6 & 32.8 & 32.9 & 44.6 & 23.0 & 41.3 & 16.5 & 23.7 & 67.7 & 38.0           \\
% ColBERT-v2          & ColBERT-base (110M) & 73.8 & 33.8 & 35.6 & 47.9 & 26.3 & 44.6 & 15.8 & 17.6 & 69.3 & 40.5           \\
ANCE           & RoBERTa-base (110M)   & 65.4 & 23.7 & 29.5 & 41.5 & 24.0 & 28.1 & 12.2 & 19.8 & 50.7 & 32.8 \\
TAS-B           & DistilBERT (66M)   & 48.1 & 31.9 & 30.0 & 42.9 & 16.2 & 38.4 & 14.9 & 22.8 & 64.3 & 34.4 \\
GenQ           & DistilBERT (66M)   & 61.9 & 31.9 & 30.8 & 49.3 & 18.2 & 32.8 & 14.3 & 17.5 & 64.4 & 35.7 \\
% GTR-11B             & - & 50.1 & 34.2 & 46.7 & 54.0 & 25.6 & 40.8 & 16.1 & 27.0 & 63.5 & 39.8 & -              \\ 
% TART-dual           & BERT-base (110M) & 62.6 & 33.7 & 33.7 & 48.9 & 20.1 & 41.5 & 14.2 & 13.8 & 69.0 & 37.4           \\ 
\hline
DPR-vanilla & Condenser-base (110M) & \textbf{69.4}  & 29.3  & \textbf{33.3}  & 41.1  & \textbf{28.0}  & \textbf{36.0}  & 14.5  & 15.6  & \textbf{54.6}  & \textbf{35.8}             \\
DPR-prompt               & Condenser-base (2.3M) & 62.8  & 30.3  & 31.5  & 40.1  & 25.1  & 33.3  & 15.2  & 15.9  & 50.3  & 33.8            \\
REMOP               & Condenser-base (2.3M×7) & 65.3  & \textbf{31.4}  & 31.6  & \textbf{41.7}  & 25.3  & \textbf{36.0}  & \textbf{15.4}  & \textbf{21.0}  & 54.0  & \textbf{35.8}            \\ 
% \hline
% DPR-vanilla  & RoBERTa-large (354M) & 58.8  & 23.2  & 35.9  & 36.8  & 25.4  & 31.2  & 13.9  & 14.0  & 47.9  & 31.9           \\
% DPR-prompt               & RoBERTa-large (6.3M) & 49.8  & 29.6  & 31.1  & 40.3  & 22.2  & 30.3  & 15.0  & 18.4  & 53.7  & 32.3           \\
% REMOP               & RoBERTa-large (6.3M×7) & 52.3  & 30.3  & 28.8  & 43.9  & 24.1  & 31.0  & 14.7  & 24.5  & 52.6  & 33.6           \\ 
\bottomrule
\end{tabular}
\renewcommand{\arraystretch}{1}
\end{adjustbox}
% The first two groups are a lexical retrieval model and sparse retrieval models. And the third group are the widely-used dense retrieval models. After that, 
\caption{Zero-shot retrieval results on BEIR, reported in NDCG@10.  \# Params are the number of trainable parameters. The last group of models is the results of DPR models based on a pre-trained condenser with different prompt settings: \textit{DPR-vanilla} (no prompts used), \textit{DPR-prompt} (adding a single prompt), and \textit{REMOP} (using modular prompts); and the highest results of the three is bolded. The labeled task attributes for each task are listed in colors: \colorbox{cblue}{Science}, \colorbox{cgreen}{QA}, \colorbox{cyellow}{Fact-Checking}, \colorbox{cred}{Wikipedia}, \colorbox{corange}{Summarization}. }
\label{table: primary experiment results}
\end{table}

% two benchmarks: BEIR (NDCG@10) and LOTTE-Search-pooled (Success@5)

As shown in Table \ref{table: primary experiment results}, compared to the fine-tuned DPR model (DPR-vanilla), REMOP can achieve comparable zero-shot retrieval performance with fewer trainable parameters. The DPR-prompt model, which is only trained on a single prefix prompt (2.3M parameters), shows relatively poor performance on zero-shot retrieval tasks; and REMOP, using module composition to enhance model generalizability, outperforms the DPR-prompt on all the evaluation tasks.
Compared to the other listed retrieval methods, REMOP shows its advantage in parameter-efficiency and generalizability that only training on a small number of parameters (2.3M×7) can achieve comparable performance. It is worth noting that most of the dense passage retrieval methods and sparse passage retrieval methods underperform the basic lexical retrieval method BM25 on zero-shot retrieval tasks, and we give an analysis of this phenomenon in Section~\ref{section: zero-shot analysis}. 

\subsection{Analysis}
\label{section: zero-shot analysis}
First, from the last group of Table \ref{table: primary experiment results}, it shows that the proposed modular retrieval approach, REMOP, outperforms the model using a single general prompt on all the evaluation tasks, which indicates that each task module (composed of several task attribute modules) can indeed match the given task and improve model specificity. Compared to the fine-tuned DPR model, REMOP demonstrates comparable performance over the evaluation tasks, which shows that it's feasible to train a few parameters and obtain fine-tuned performance on zero-shot retrieval scenarios.

Second, it is also clear to see that the results of most of the listed dense passage retrieval methods are lower than the BM25, which is an acceptable phenomenon since previous research \cite{ren2022thorough} has indicated that dense passage retriever perform worse than lexical retrieval models on zero-shot retrieval task. From this perspective, we can consider the proposed modular retrieval paradigm as a feasible solution to alleviate this shortcoming of dense retrieval. 

Lastly, besides parameter efficiency and generalization, the interpretability advantage of REMOP is also shown in the results. For example, REMOP can have a higher enhancement over the fine-tuned model and the single-prompt model on Fact-Checking-related tasks, which indicates that the Fact-Checking retrieval module is well-trained in the training phase. This observation provides a valuable reference for subsequent improvements, guiding us in which part we should invest more training resources for improvement.

\section{Related Works on Modularity}
\label{section: related works}

Modular learning techniques have been extensively utilized in machine learning and demonstrated impressive performance in various tasks such as machine translation~\cite{bapna2019simple,nllbteam2022language}, speech processing~\cite{le2021lightweight}, and computer vision~\cite{he2022parameter}. DEMIX~\cite{gururangan2021demix} employs expert layers as modules to store domain-specific knowledge and leverages modularity to enhance model generalization. 
ATTEMPT~\cite{asai2022attentional} trains task-specific soft prompts for each task and combines them using attention routing, which is a one-to-one mapping case discussed in Section~\ref{section: task decomposition}. 
MAD-X~\cite{pfeiffer2020mad} and AdapterSoup~\cite{chronopoulou2023adaptersoup} employ adapter structures~\cite{pfeiffer2020mad,Rebuffi_2018_CVPR} to implement domain-specific modules \footnote{Compared to other parameter-efficient methods, adapter layers usually require more parameters as they depend on a model's input and hidden size for the new function~\cite{pfeiffer2023modular}.}. The former uses adapter layers with fixed routing for zero-shot cross-lingual transfer, while the latter employs a text clustering approach for module routing.

As shown in Table \ref{table: related works comparisons}, compared with the previous works, the main differences of our proposed approach are: 1) we first apply modular learning in the \textbf{information retrieval} field; 2) we enhance the \textbf{generalizability} of modularity by using task attribute decomposition to implement many-to-many task module mapping; 3) we associate modules with specific task attributes, which vastly improves the \textbf{interpretability} of modularity; 4) lastly, we implement the module using a \textbf{parameter-efficient approach}, which is friendly to low-resource scenarios and efficient in computation and storage.

\begin{table}[htp]
\centering
\begin{adjustbox}{width=\columnwidth,center}
\begin{tabular}{lcccc}
\toprule
\textbf{Related Works} & Information Retrieval & Generalizability & Interpretability & Parameter-efficiency \\ \midrule
\rowcolor{lightgray} \textbf{REMOP}      & $\surd $    & $\surd $    & $\surd $    & $\surd \surd $              \\
DPR~\cite{karpukhin2020dense}               &  $\surd $  &  &  &               \\
\rowcolor{cgray} DPTDR~\cite{tang2022dptdr}               &  $\surd $  &  &  &  $\surd \surd $               \\
DEMIX~\cite{gururangan2021demix}               &    & $\surd $ &  $\surd $  &               \\
\rowcolor{cgray} ATTEMPT~\cite{asai2022attentional}             &    & $\surd $  &    & $\surd \surd $    \\
MAD-X~\cite{pfeiffer2020mad}               &    &   & $\surd $    & $\surd $    \\
\rowcolor{cgray} AdapterSoup~\cite{chronopoulou2023adaptersoup}         &    & $\surd$   & $\surd$ & $\surd$ \\
\bottomrule
\end{tabular}
\end{adjustbox}
\caption{Comparison with Related Works.}
\label{table: related works comparisons}
\end{table}

\section{Conclusion}
\label{conclusion}
This paper formally defines a new information retrieval paradigm, modular retrieval, which aims to bring the benefits of modular learning to information retrieval tasks and increase the interpretability of the retrieval process. Based on the proposed retrieval paradigm, we propose a novel modular retrieval method REMOP, which utilizes three modular operations to flexibly combine retrieval modules to perform zero-shot retrieval tasks, with high interpretability and generalization ability. We explore and verify the effectiveness of the proposed module arithmetic, and experiments on a zero-shot retrieval benchmark show that REMOP has comparable performance to the fine-tuned model in zero-shot retrieval.

% \section*{Acknowledgements}
% similarities and differences between the related works and the this work

% \section*{References}
{
\small
\bibliographystyle{unsrt}
\bibliography{main}
}

% References follow the acknowledgments. Use unnumbered first-level heading for
% the references. Any choice of citation style is acceptable as long as you are
% consistent. It is permissible to reduce the font size to \verb+small+ (9 point)
% when listing the references.
% Note that the Reference section does not count towards the page limit.
\clearpage

\appendix

\end{document}